\documentclass[
]{ceurart}

\usepackage[english]{babel}    %
\usepackage{listings}
\usepackage[utf8]{inputenc}

\usepackage{float}
\usepackage{graphicx}
\usepackage{pdflscape}

\begin{document}

\copyrightyear{2023}
\copyrightclause{Copyright for this paper by its authors.
  Use permitted under Creative Commons License Attribution 4.0
  International (CC BY 4.0).}

\conference{ISWC 2023 Posters and Demos: 22nd International Semantic Web Conference, November  6–10, 2023, Athens, Greece}

\title{Linked Papers With Code: The Latest in Machine Learning as an RDF Knowledge Graph}

\author{Michael Färber}[%
orcid=0000-0001-5458-8645,
email=michael.faerber@kit.edu,
]
\cormark[1]
\address{Karlsruhe Institute of Technology (KIT), Institute AIFB, Germany}

\author{David Lamprecht}[%
orcid=0000-0002-9098-5389,
email=david.lamprecht@student.kit.edu,
]

\cortext[1]{Corresponding author.}

\begin{abstract}
In this paper, we introduce \textit{Linked Papers With Code} (LPWC), an RDF knowledge graph that provides comprehensive, current information about almost 400,000 machine learning publications. 
This includes the tasks addressed, the datasets utilized, the methods implemented, and the evaluations conducted, along with their results. 
Compared to its non-RDF-based counterpart \textit{Papers With Code}, LPWC not only translates the latest advancements in machine learning into RDF format, but also enables novel ways for scientific impact quantification and scholarly key content recommendation. LPWC is openly accessible at \url{https://linkedpaperswithcode.com} and is licensed under CC-BY-SA 4.0. As a knowledge graph in the Linked Open Data cloud, we offer LPWC in multiple formats, from RDF dump files to a SPARQL endpoint for direct web queries, as well as a data source with resolvable URIs and links to the data sources SemOpenAlex, Wikidata, and DBLP. 
Additionally, we supply knowledge graph embeddings,
enabling LPWC to be readily applied in machine learning applications. 
\end{abstract}


\maketitle

\section{Introduction}

Over the years, several scientific knowledge graphs have emerged, including ORKG~\cite{ORKG2020}, MAKG~\cite{farber_makg_2019}, and most recently, SemOpenAlex~\cite{semopenalex}. 
However, no knowledge graph exists that explicitly targets the modeling of key content in machine learning on a large scale and always up-to-date. On the other hand, \textit{Papers with Code} (PWC,  \url{https://paperswithcode.com}) has emerged as a platform for machine learning publications, code, datasets, methods, and evaluation tables that can be updated by anyone. However, PWC is only available as a web page and a JSON dump without semantic modeling and Linked Open Data (LOD) integration. 

We introduce \textit{Linked Papers With Code} (LPWC) -- 
the first RDF knowledge graph that comprehensively models the research field of machine learning using an extensive ontology. Our knowledge graph goes beyond simply lifting to RDF format, for example, by resolving complex data formats through graph modeling and linking entities to other LOD sources such as SemOpenAlex, Wikidata, and DBLP.
LPWC consists of 7,935,279 RDF triples as of June 2023 and is available at \url{https://linkedpaperswithcode.com}. 
It has multiple applications, from improved management of research data to more effective integration of data across different research domains. By incorporating FAIR principles that focus on the availability and reuse of research data and artifacts, we expect LPWC to 
improve the discoverability and applicability of machine learning research results. We make the code used for knowledge graph creation and embedding generation available 
online (\url{https://github.com/davidlamprecht/linkedpaperswithcode}). In the following, we present LPWC in detail.

\section{Linked Papers with Code}

\textbf{Linked Papers With Code Ontology.} First, we develop an ontology that adheres to the best practices of ontology engineering and incorporates as much existing vocabulary as possible. 
Given that the PWC data dump is sourced directly from the PWC website, thus lacks a standardized schema and comprises diverse JSON objects, it was infeasible to directly model it within an OWL/RDF framework. Consequently, 
we construct a novel semantic schema to model the data. 
An overview of the entity types, object properties, and data type properties can be found in Figure~1. The LPWC ontology encompasses 
\textit{13 entity types} and \textit{47 relationship types}. 
In addition to the ontology, which is available as an OWL file, 
we provide a VoID file, following the Linked Open Data good practices to describe our linked dataset.

\textbf{Linked Papers With Code Knowledge Graph. }
PWC 
provides access to its data via a user-friendly, human-readable website. In addition, it offers daily JSON data dumps.\footnote{See \url{https://github.com/paperswithcode/paperswithcode-data}} 
However, there are several aspects that currently make using the data difficult:
1.~There is a lack of semantic interoperability. Entities, such as authors or AI models, are represented as strings without unique IDs. This prevents effective linking of data and creation of knowledge graphs.
2. %
Due to the complexity of the data, modeling in JSON format proves difficult, especially when processing or querying the data. This issue becomes particularly apparent with evaluation tables, which are nested within a JSON structure with up to 19 levels in depth. This results in significant data redundancy within the file. In contrast, a graph representation provides a more intuitive and manageable way of modeling.
3.  The data, originally designed for a human readable interface, uses markdown for natural language descriptions of entities, which may not be optimal when being processed by NLP methods or displaying it outside of the website.

\textit{Data Transformation. } 
To overcome these limitations, we convert the 
JSON files from the PWC data dump into an RDF knowledge graph based on the developed ontology. This requires major changes in the data formatting and data modeling.
In the transformation process we, among other steps, (1)~assign unique HTTP URIs to all entities, (2)~convert all markdown test to plain text and (3)~link the entities to other scholarly data sources in the LOD cloud.

\textit{Author Name Disambiguation. } The disambiguation of author names given as strings is a crucial step on top of the pure data transformation. 
Specifically, we develop an efficient two-step method to link the 1,471,006 authors 
in LPWC to 
entities in SemOpenAlex, 
which is a massive RDF dataset modeling the academic landscape with its publications, authors, sources, and institutions, via its public SPARQL endpoint \cite{semopenalex}. 
We leverage LPWC author names and paper titles for the disambiguation. 
The first step involves exact name matching and publication title substring comparison.

\begin{landscape}
\centering 
\begin{figure}
    \centering
    \includegraphics[width=0.95\linewidth]{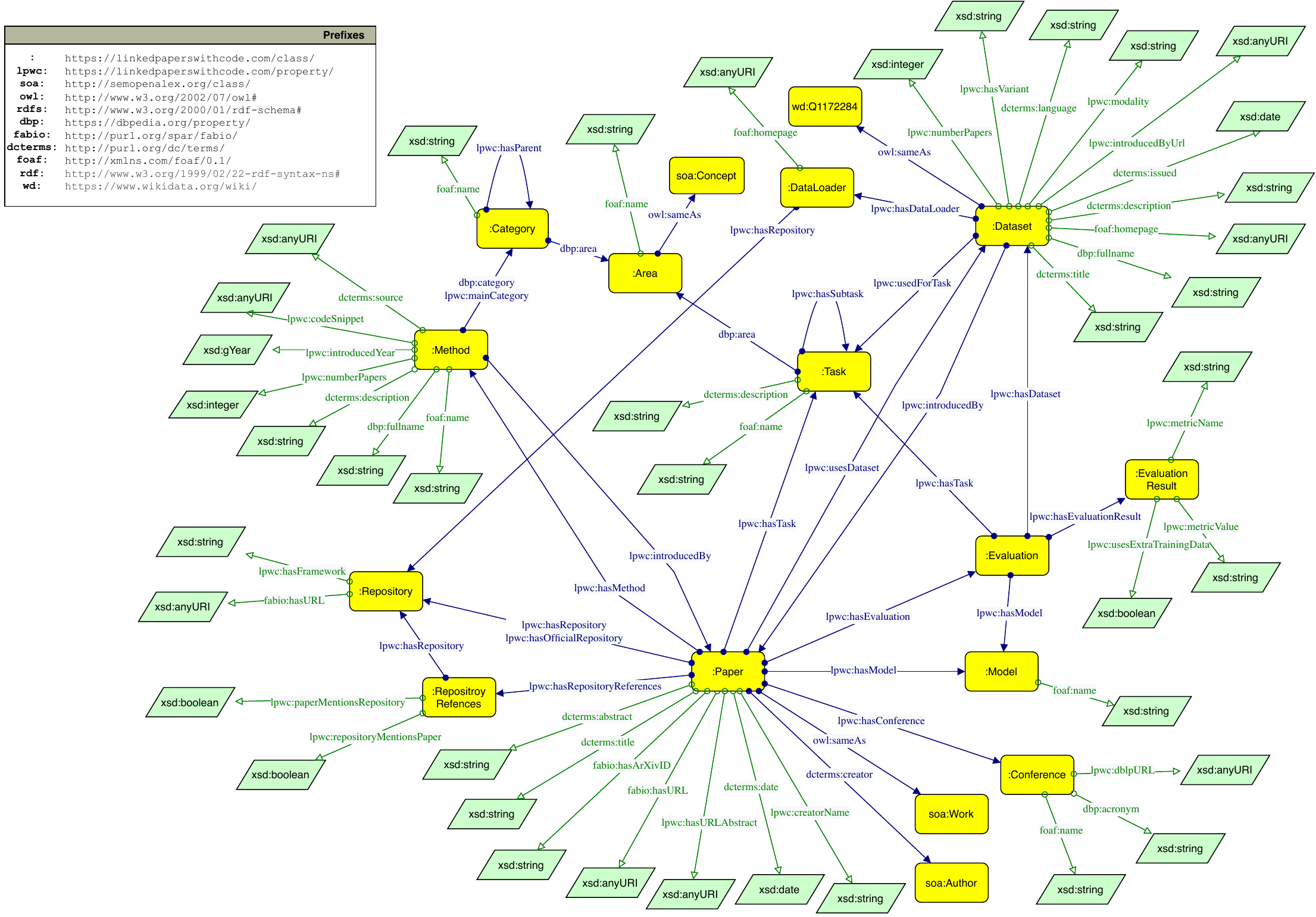} %
        \caption{\textit{Schema of Linked Papers With Code}.} 
        \label{fig:KGSchema}
\end{figure}
\end{landscape}

If no match is found, the second step employs a variant search of LPWC paper titles in SemOpenAlex works, 
and author matching based on fuzzy similarity techniques.
This process yields 947,709 links to SemOpenAlex entities. 
The remaining 523,297 author names 
are 
represented in LPWC using the \texttt{lpwc:authorName} property.

\textit{Creating owl:sameAs statements. }We further link all conferences modeled in LPWC to DBLP.
Moreover, we successfully map 267,314 papers (71\% of all papers in LPWC) to SemOpenAlex works, utilizing variations of the LPWC paper titles. 
Lastly, we are able to create 158 mappings (2\% of all datasets) between datasets modeled in LPWC and datasets modeled in Wikidata.

\textbf{Key Statistics. }
Our knowledge graph's SPARQL endpoint enables the direct computation of interesting statistics. For instance, Table 1 shows the frequency of entities across entity types. Additionally, Figure 2 illustrates how to compare conferences (here: NAACL, EMNLP, ACL) based on the used evaluation metrics of their papers.

\begin{figure}[tb]
  \begin{minipage}{0.33\textwidth}
    \begin{center}
    \begin{tabular}{lr}
      \toprule
      \textbf{Entity Type} & \textbf{\# Instances} \\ 
      \midrule 
         Paper    & 376,557 \\
         Evaluation & 52,519 \\
         Paper with Evaluations    & 13,289 \\
         Repository    &  153,476 \\
         Model    & 24,598  \\
         Dataset    & 8,322  \\
         Task    & 4,267  \\
         Method    & 2,101  \\
         Conference    & 1,407  \\
      \bottomrule
    \end{tabular}
    \captionof{table}{Linked Papers With Code entity types \\ and number of instances (as of June 2023).}
    \label{fig:table1}
    \end{center}
  \end{minipage}
  \hfill
  \begin{minipage}{0.5\textwidth}
    \centering
    \includegraphics[width=0.82\textwidth]{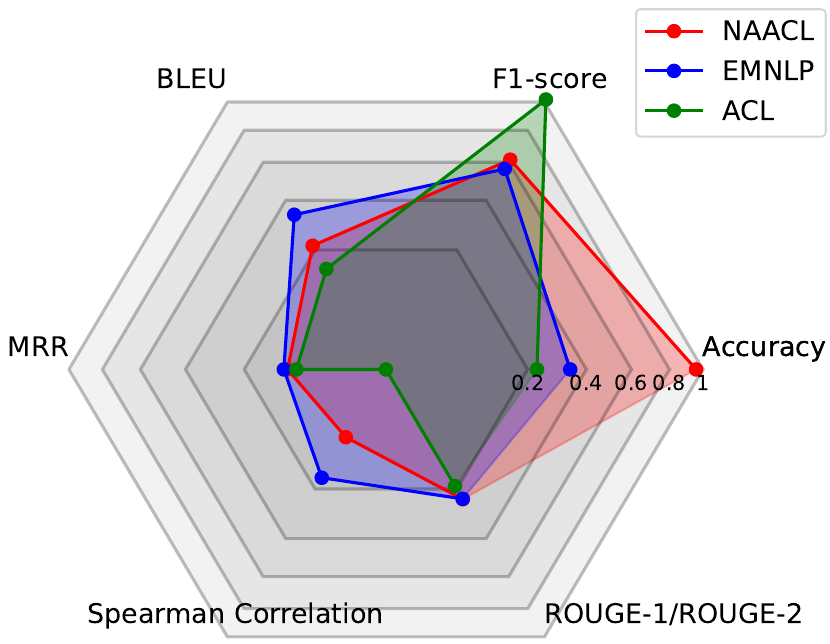}
    \caption{Distribution of evaluation metrics\\ used in NAACL, EMNLP, and ACL conferences.}
    \label{fig:graph1}
  \end{minipage}
\end{figure}

\textbf{Knowledge Graph Embeddings. } 
To enable additional use cases, we compute knowledge graph embeddings for LPWC.
Embeddings have proven to be valuable as implicit knowledge representations in various scenarios. We train the embeddings based on state-of-the-art embedding techniques such as TransE, DistMult, ComplEx, and RotatE \cite{wang2019knowledge,demir2021convolutional}. The training process involves a maximum of 900 epochs, implementing early stopping based on the mean rank calculated on the validation sets at intervals of 300 epochs. 
Among the evaluated techniques, TransE shows the best results. Therefore, we provide the TransE-based embedding vectors for all entities and relations online 
and all our evaluation results in our repository.
Notably, our provided embeddings are in line with state-of-the-art results on benchmark datasets with similar characteristics in terms of the number of relations, triples, and entities \cite{wang2019knowledge,demir2021convolutional}.

\textbf{Use Case Examples. }
LPWC can enhance existing use cases while also enabling the development of new ones. In the following, we highlight some potential use cases: %
\begin{enumerate}
 \item \textit{Machine Learning Data Analysis:} LPWC is a novel scientific knowledge graph 
covering the current field of machine learning. 
Complex analyses, such as comparing conferences or detecting new research topics, become possible in this way. 
\item \textit{Scholarly LOD Cloud Enrichment:}
LPWC is highly integrated with the LOD cloud and connected to multiple data sources such as  SemOpenAlex, Wikidata, and DBLP.
This enables efficient data integration and enhanced research data management 
in alignment with 
the FAIR principles.  
\item \textit{Academic Recommender Systems:}
Given the information overload in science, scientific recommender systems are becoming increasingly important. %
LPWC and the provide knowledge graph embeddings can be used directly to build state-of-the-art recommender systems for key scientific content. With LPWC, these systems can recommend also items such as datasets, methods, and conferences. %
\end{enumerate}

\vspace{-0.15cm}
\section{Conclusion}
\vspace{-0.05cm}
In this paper, we presented \textit{Linked Papers with Code}, the first RDF knowledge graph with detailed information about the machine learning landscape, consisting of close to 8 million RDF triples. We outlined the creation process of this dataset, discussed its characteristics, and examined the procedure for training state-of-the-art knowledge graph embeddings. 
In future work, we aim to leverage the extensive interconnectivity between LPWC and SemOpenAlex to facilitate large-scale key content extraction from publications.

\vspace{-0.3cm}
\bibliography{bibliography}

\appendix

\end{document}